\documentstyle[sprocl]{article}

\input{psfig}

\def\Journal#1#2#3#4{{#1} {\bf #2}, #3 (#4)}

\def\be{\begin{equation}}
\def\ee{\end{equation}}
\def\bea{\begin{eqnarray}}
\def\eea{\end{eqnarray}}
\newcommand \ga{\raisebox{-.5ex}{$\stackrel{>}{\sim}$}}
\newcommand \la{\raisebox{-.5ex}{$\stackrel{<}{\sim}$}}

\begin{document}

\title{SIGNATURES OF QCD MATTER AT RHIC}

\author{H. HEISELBERG}

\address{NORDITA, Blegdamsvej 17\\ DK-2100 Copenhagen \O., DENMARK\\E-mail: 
hh@nordita.dk} 

\author{A. D. JACKSON}

\address{Niels Bohr Institute, Blegdamsvej 17\\ DK-2100 Copenhagen \O., 
DENMARK\\E-mail: 
jackson@nbi.dk} 


\maketitle\abstracts{We discuss possible experimental
signatures of forming a Quark-Gluon plasma in high
energy nuclear collisions. In first order phase
transitions such as the chiral symmetry restoration supercooling
may lead to density fluctuations such as droplet formation and
hot spots.  These will lead to rapidity fluctuations event-by-event
and triggering on such fluctuations will have distinct effects for
HBT radii.}

\section{Introduction}

Results from the RHIC collider next year are eagerly awaited. The hope
is to observe the phase transition to
quark-gluon plasma, the chirally restored hadronic matter and/or
deconfinement. This may be by distinct signals of J/$\Psi$ suppression,
strangeness enhancement, $\eta'$ enhancement, enhanced temperature
and multiplicity fluctuations \cite{Shuryak}, etc.
Other signals are plateau's in temperatures, transverse flow or other
collective quantities as function of centrality, transverse energy or
multiplicity.
 We will here suggest other signals of a first order phase transition
namely rapidity fluctuations and associated variations in HBT radii.

\section{Supercooling and Droplet formation}

Lattice calculations suggest that QCD has a first order transition at 
zero chemical potential provided that the strange quark is sufficiently 
light \cite{Lattice}. 
Recent work using random matrix theory (RMT) \cite{Jackson} suggests 
an effective thermodynamic potential of the form
\begin{equation}
\Omega(\phi) /N_f = \phi^2 - \frac{1}{2}{\rm ln} \{ [ \phi^2 - 
(\mu + iT)^2] \cdot [\phi^2 - (\mu - iT)^2] \} \ .  \label{Omega}
\end{equation}
The value of $\phi \sim \langle {\bar \psi} \psi \rangle$ at the 
minima of $\Omega (\phi)$ is related to the expectation value of the quark 
condensate which is the order parameter for chiral symmetry breaking.  
Minimization of Eq.(\ref{Omega}) leads to a fifth order polynomial equation 
for $\phi$ which is identical in form to the results of Landau-Ginzberg 
theory using a $\phi^6$ potential.  One solution to this equation corresponds 
to the restored symmetry phase with $\phi = 0$.  This model predicts 
a second order transition for $\mu = 0$ at a temperature, $T_c$, which 
is generally agreed to be approximately 140 Mev.  For $T=0$, a first 
order transition occurs at some $\mu_0$.  Since the phases in 
which chiral symmetry is broken and restored must be separated in the 
$( \mu , T )$ plane by an unbroken line of phase transitions, this 
implies the existence of a tricritical point.  In this model, the 
tricritical point is at 
\begin{displaymath}
\frac{T_3}{T_c} = \frac{1}{2}\sqrt{\sqrt{2}+1} \ \ {\rm and} \ \ 
\frac{\mu_3}{\mu_0} \approx 0.610 \ .
\end{displaymath}
Further, the discontinuity in the 
density at $(T=0 , \mu_0)$ can be estimated as $\Delta n \approx 
2.5 n_0$ where $n_0$ is the equilibrium density of nuclear matter.  An 
inevitable consequence of such a phase diagram is the existence of 
spinodal lines within which $\Omega (\phi )$ has three local minima.  
In this region, nuclear matter can be either superheated or super cooled.  

The qualitative features of this phase diagram lead to an interesting 
scenario for relativistic heavy ion collisions in which matter is 
compressed and heated.  Some or all of this matter will undergo chiral 
restoration.  If the subsequent expansion is sufficiently rapid, matter 
will pass the phase coexistence curve with little effect since the 
restored symmetry phase with $\phi = 0$ will remain a local minimum of 
$\Omega ( \phi )$.  This suggests the possible formation of ``droplets'' 
of supercooled chiral symmetric matter with relatively high baryon and 
energy densities in a background of low density broken symmetry matter.  
These droplets can persist until the system reachs the spinodal line and 
then return rapidly to the now-unique broken symmetry minimum of $\Omega 
( \phi )$.  The schematic nature of Eq.(\ref{Omega}) makes it impossible to 
make quantitative predictions about the properties of such droplets.  
(The effective thermodynamic potential can well have additional terms 
which are independent of $\phi$.)  However, a large mismatch in density 
and energy density seems to be a robust prediction.  Note that these 
``droplets'' are kinematically distinct from the ``hot spots'' usually 
associated with jets and minijets \cite{Gyulassy}.  
Jets are characterized by a 
strong directional orientation and a high transverse momentum not expected 
for the present droplets.

Such density fluctuations can occur only for first order transitions.  
In the RMT model, these will only be the case for sufficiently high 
net baryon densities.  In ultrarelativistic nuclear collisions at 
mid rapidity, the baryon density is low as in the early universe and, 
according to RMT, the transition is of second order.  One should then 
investigate nuclear fragmentation regions or go to lower collision 
energies (e.g., AGS energies) to observe droplet formation.  However, 
most lattice simulations do find a first order transition at zero baryon 
density.

\section{Rapidity Fluctuations}

 If the transitions is first order, matter may supercool and
subsequently create fluctuations in a number of quantities.  Density
fluctuations in the form of hot spots or droplets of dense matter with
hadronic gas in between is a likely outcome.  We shall refer to these
regions of dense and hot matter in space-time as well as in momentum
space as droplets.
If we assume that hadrons emerge as 
a Boltzmann distribution with temperature $T$ from each droplet and
ignore transverse flow, the 
resulting particle distribution is
\be
 \frac{dN}{dyd^2p_\perp} \propto  \sum_i f_i\, e^{-m_\perp\cosh(y-\eta_i)/T} 
     \,. \label{dNdy}
\ee
Here, $y$ is the particle rapidity and $p_\perp$ its transverse momentum,  
$f_i$ is the number of particles
hadronizing from each droplet {\it i}, and  
\be
\eta_i=\frac{1}{2}\log\frac{t_i+z_i}{t_i-z_i}=\frac{1}{2}\log
 \frac{1+v_i}{1-v_i}
  \, 
\ee
is the rapidity of droplet {\it i}.

When $m_\perp/T\gg 1$, we can approximate
$\cosh(y-\eta_i)\simeq 1+\frac{1}{2}(y-\eta_i)^2$ in Eq.(\ref{dNdy}). 
The Boltzmann factor determines the
width of the droplet rapidity distribution as $\sim \sqrt{T/m_\perp}$. 
The rapidity distribution will display fluctuations
in rapidity event by event
when the droplets are separated by rapidities larger than
$|\eta_i-\eta_j|\ga \sqrt{T/m_\perp}$. If they are evenly distributed
by smaller rapidity differences, the resulting rapidity distribution
(\ref{dNdy}) will appear flat.

The droplets are separated in rapidity by 
$|\eta_i-\eta_j|\sim \Delta z/\tau_0$, where $\Delta z$ is the
correlation length in the dense and hot mixed phase and $\tau_0$ is 
the invariant time after collision at which the droplets form. Assuming
that $\Delta z\sim 1$fm --- 
a typical hadronic scale --- and that the droplets form
very early $\tau_0\la 1$fm/c, we find that indeed
$|\eta_i-\eta_j|\ga \sqrt{T/m_\perp}$ even for the light pions. 

If strong transverse flow is present in the source, the droplets may also
move in a transverse direction. In that case the distribution in 
$p_\perp$ may be 
non-thermal and azimuthally asymmetric.

\begin{figure}[t]
\vskip -.5cm
\psfig{figure=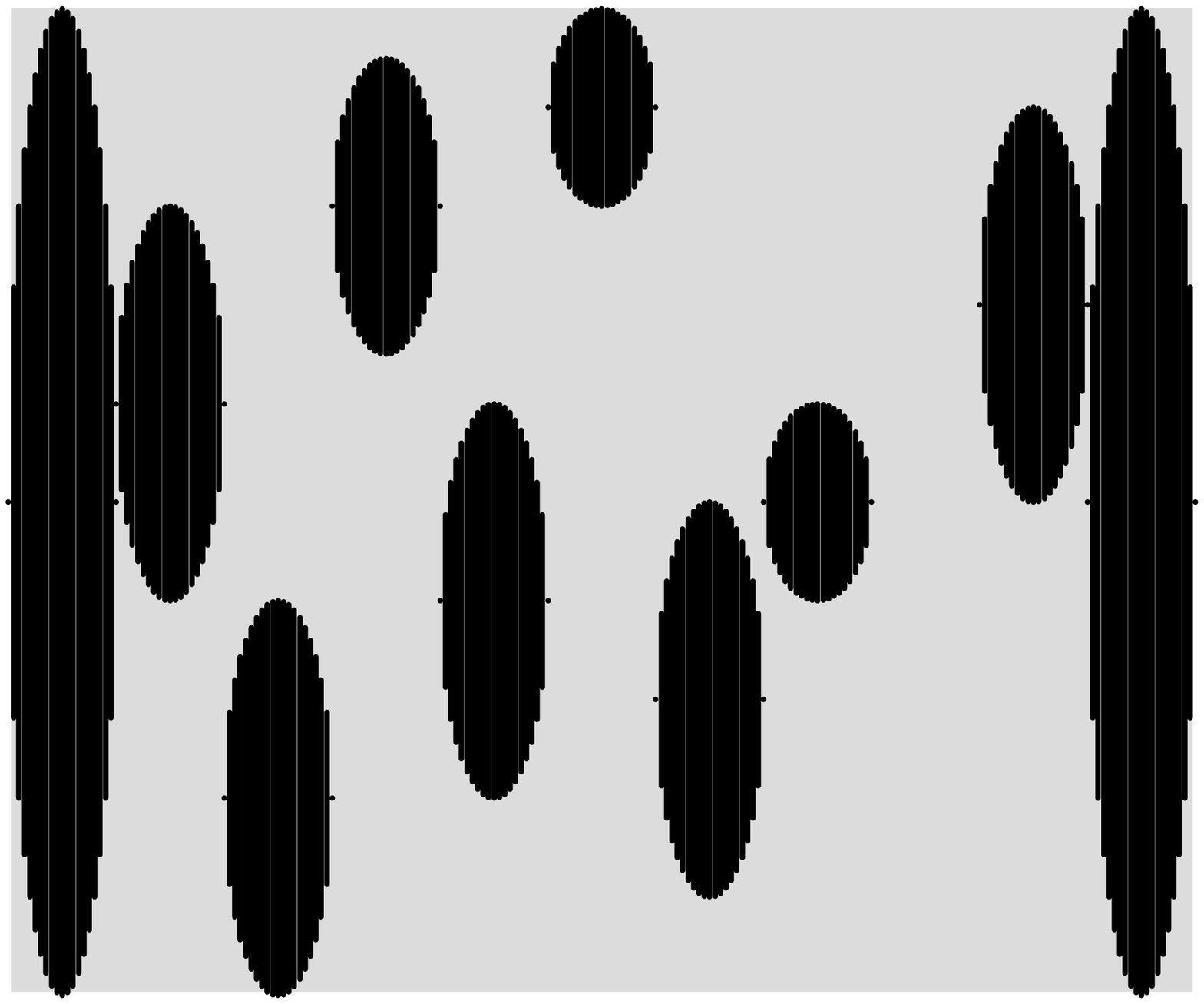,height=2.5in,angle=-90}
\vspace{-1.5cm}
\psfig{figure=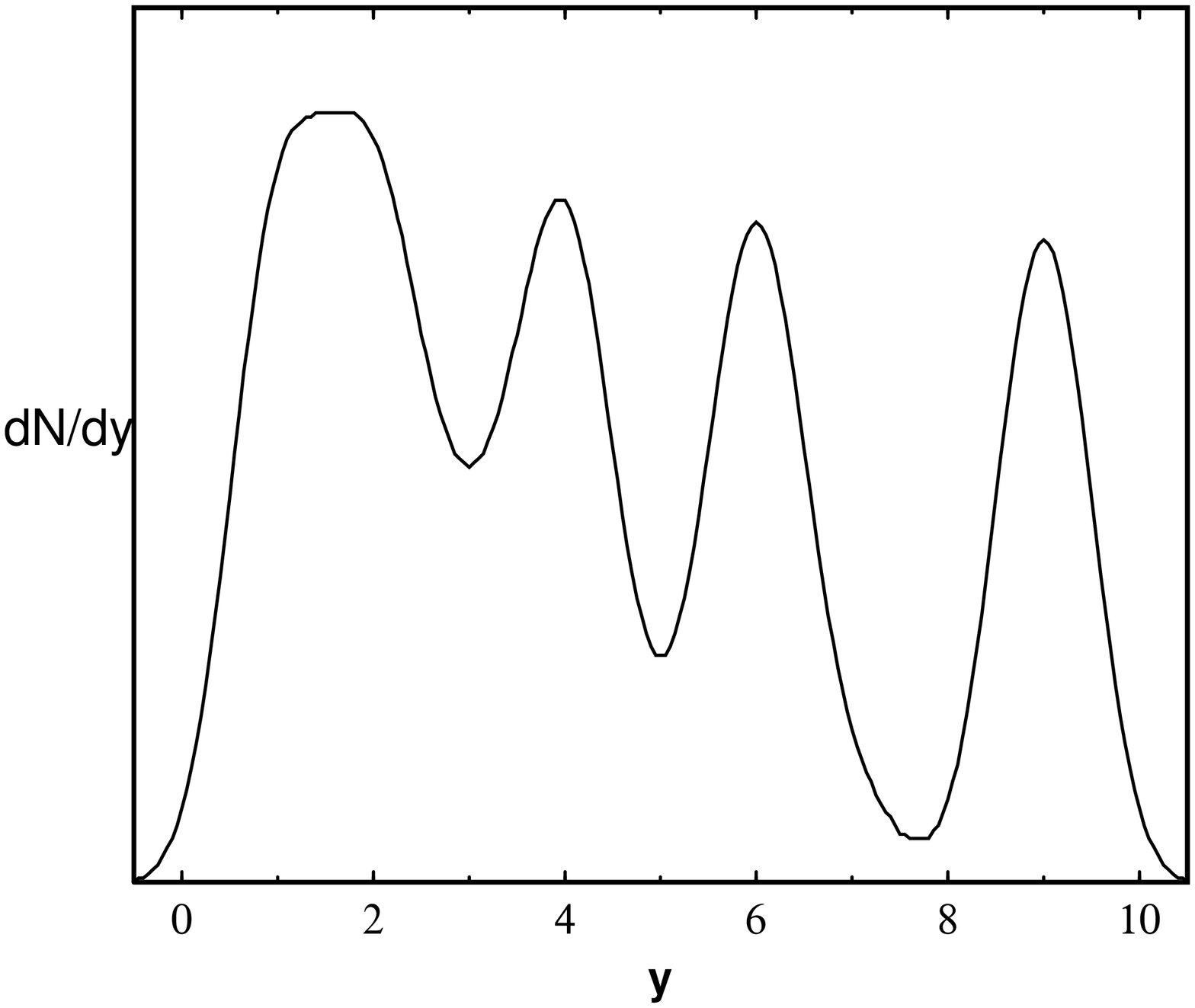,height=2.5in,angle=-90}
\caption{Droplet formation (top) and corresponding
rapidity distribution (bottom).  \label{Fig1}}
\end{figure}

\newpage

\section{HBT and Correlation Functions}

In this section we briefly review 
the correlation function analysis of Bose-Einstein interference.
For details we refer to references on stellar intensity interferometry
\cite{HBT}, pion interferometry in pp \cite{pp} and AA collisions
\cite{GKW,Csorgo,Heinz}.

For a source of size $R$ we consider two particles emitted a distance
$\sim R$ apart with relative momentum ${\bf q}=({\bf k}_1-{\bf k}_2)$
and average momentum, ${\bf K}=({\bf k}_1+{\bf k}_2)/2$. Typical heavy
ion sources in nuclear collisions are of size $R\sim5$ fm, so that
interference occurs predominantly when 
$q\raisebox{-.5ex}{$\stackrel{<}{\sim}$}\hbar/R\sim 40$
MeV/c. Since typical particle momenta are $k_i\ga K\sim 300$ MeV/c,
the interfering particles travel almost parallel, i.e.,
$k_1\simeq k_2\simeq K\gg q$.  The correlation function due to
Bose-Einstein interference of identical particles from an incoherent
source is (see, e.g., \cite{Heinz})
\begin{equation}
  C_2({\bf q},{\bf K})=1\;\pm\; \left|\frac{\int d^4x\;S(x,{\bf K})\;e^{iqx}}
   {\int d^4x\;S(x,{\bf K})}\right|^2 \,, \label{C}
\end{equation}
where $S(x,{\bf K})$ is the source distribution 
function describing the phase space density of the
emitting source. The $+/-$ refers to boson/fermions respectively.

Experimentally the correlation functions for identical mesons
($\pi^\pm\pi^\pm$,\\ 
$K^\pm K^\pm$, etc.) are often
parametrized by the gaussian form
\bea
  C_2(q_s,q_o,q_l)=1+\lambda\exp\left[
           - q_s^2R_s^2-q_o^2R_o^2-q_l^2R_l^2
            -2q_oq_lR_{ol}^2  \right]\;.   \label{Cexp}
\eea
Here, ${\bf q}={\bf k}_1-{\bf k}_2=(q_s,q_o,q_l)$ is the relative momentum
between the two particles and $R_i,i=s,o,l$ the corresponding sideward,
outward and longitudinal HBT radii respectively. We have suppressed the
${\bf K}$ dependence.
We will employ the standard geometry, where the {\it longitudinal} direction
is along the beam axis, the {\it outward} direction is along ${\bf K}$, and
the {\it sideward} axis is perpendicular to these.
Usually, each pair of mesons is lorentz boosted longitudinal to the system 
in which their rapidity vanishes, $Y=0$. Their 
average momentum
${\bf K}$ is then perpendicular to the beam axis and is chosen as the outward
direction. In this system the pair velocity
\mbox{\boldmath $\beta_K$}=${\bf K}/E_K$ points in
the outward direction with $\beta_o=p_\perp/m_\perp$ where 
$m_\perp=\sqrt{m^2+p_\perp^2}$ is the transverse mass.
Often models assume a boost invariant source and boost to the cms where $Y=0$
where a number of boost factors disappear and 
the out-longitudinal coupling $R_{ol}$ vanishes.
The reduction factor $\lambda$ in Eq.(\ref{Cexp}) may be
due to long lived resonances \cite{Csorgo,HH}, coherence effects,
incorrect Gamow corrections \cite{BB} or other effects. It is found to
be $\lambda\sim 0.5$ for pions and $\lambda\sim 0.9$ for kaons.

\section{HBT for Droplets}

Droplets lead to spatial fluctuations in density which can be
probed by correlations between identical particles \cite{HBT}.
For simplicity we parametrize these droplets by spatial and temporal
gaussians of
size $R_d$ and duration of emission $t_d$. The distribution of
particles in space and time --- usually referred to as the source ---
for droplets situated at $x_i=({\bf r}_i,t_i)$ is
\be
 S(x,K)\sim \sum_i \tilde{S}(x_i,K) 
             \exp\left[ -\frac{( {\bf r-r}_i)^2}{2R_d^2} 
                         -\frac{(t-t_i)^2}{2t_d^2} \right] \,, \label{SD}
\ee
where $\tilde{S}(x_i,K)$ is the distribution of sources.
Normalizations of $S$ cancel when calculating correlation functions, 
Eq.(\ref{C}). Other
effects as transverse flow, opacities \cite{Opaque}, resonances
\cite{HH}, etc.  can be included but we shall ignore
them here for simplicity.
Likewise, the extension to droplets of different size and duration of emission
is straight forward but unnecessary for our purpose.
We expect the scale of  $\tilde{S}$ is the nuclear overlap
size  $R_A$ of order several $fermi$'s whereas
the droplet size $R_d$ is smaller --- of order a few $fermi$'s.

From (\ref{C}) and (\ref{SD}) we obtain
\be
 C_2(q) \simeq 1 + \exp[-{\bf q}^2R_d^2
       -({\bf q\cdot\beta})^2t_d^2]\, |\tilde{S}(q,K)|^2 \,, \label{CD}
\ee
where
\be
  \tilde{S}(q,K)=\sum_i \tilde{S}(x_i,K) 
 e^{iq\cdot x_i}/\sum_i \tilde{S}(x_i,K)
\ee
is the Fourier transform of the distribution of sources and
 ${\bf\beta}={\bf K}/K_0$ is the average velocity of the pair.

If there is only a single droplet $\tilde{S}(q,K)=1$ and the
correlation function is simply given by the droplet gaussian.   
If there are many droplets, the sum
can be replaced by an integral.

\subsection{Non-expanding Sources}

For non-expanding sources we can assume that the particle momentum 
distribution is the same for all droplets, i.e., independent of ${\bf K}$.

For many droplets we assume that the droplets are distributed by a 
gaussian 
\be
 \tilde{S}(x_i,K)\sim\exp(-{\bf r}_i^2/2R_A^2-t_i^2/2R_A^2) \,,
\ee
where $t_A$ is the spread in droplet formation
times and  $R_A$ is the size of the nuclear overlap zone.
It increases with decreasing impact parameter - roughly as
$R_A(b) \simeq \sqrt{R_A(0)^2-b^2}$, where $R_A(0)\simeq 1.2 A^{1/3}fm$
is the nuclear size.

The resulting correlation function becomes
\be
 C_2(q) \simeq 1 + \exp[-{\bf q}^2(R_A^2+R_d^2)
                  -({\bf q\cdot\beta})^2(t_A^2+t_d^2)] 
                  \,. \label{CA}
\ee
Since $R_A$ and $ct_A$ are typically of nuclear scales $\sim 5-10$ fm
whereas we expect $R_d\sim 1$ fm, the droplets are simply ``drowned''
in the background of the other droplets.

In the case of only two droplets (of same size) the resulting
correlation function is
\be
 C_2(q) \simeq 1 + \exp[-{\bf q}^2R_d^2
                   -({\bf q\cdot\beta})^2t_d^2] \,
                   \cos(\frac{1}{2}q\cdot(x_1-x_2)) \,. \label{Ccos}
\ee  
However, as the $x_i$'s differ from event to event
the oscillation in (\ref{Ccos}) will, when summing over events,
result in a Fourier transform over the
distribution of $x_i$'s in different event and the end result will be
similar to (\ref{CA}).
The number of pairs in a single event at RHIC energies 
is probably not large enough
to do event-by-event HBT. The oscillation will also be smeared by resonances, 
Coulomb effects, a hadronic background, etc.

We conclude that it will be very difficult to see the individual droplets
through HBT when the sources do not expand.

\subsection{Expanding Sources}

At RHIC energies the collision regions are rapidly expanding particularly
in the longitudinal direction. Consequently the
droplets may have different expansion velocities and rapidities and
the distribution $\tilde{S}(x_i,K)$ {\it  will 
depend on $K$} and this difference will now be exploited.

When the droplets are separated in rapidity such that 
$|\eta_i-\eta_j|\ga\sqrt{T/m_\perp}$, only particles within the same droplet
contribute to the correlation function.
If we transform to the droplet center-of-mass system, i.e. $\eta_i=0$,
the correlation function is then given by (\ref{CA}) with $\tilde{S}=1$.
In terms of $q_s,q_o,q_l$ it can be written
\bea
 C_2(q)&=&1+\exp\left[ -q_s^2R_d^2 -q_o^2(R_d^2+\beta_o^2t_d^2)
         -q_l^2(R_d^2+\beta_l^2t_d^2) -2q_oq_l\beta_o\beta_l t_d^2 \right]
   \,, \nonumber \\
  &&  \label{Cdy}
\eea
where $\beta_o=p_\perp/m_\perp\cosh(Y)$ and $\beta_l=\tanh(Y)$, are the
transverse (outward) and longitudinal velocities respectively
of the pair.

When the droplets overlap in rapidity, we need to consider the distribution
of droplets in more detail.
At ultrarelativistic energies the rapidity distribution is expected to
display approximate Bjorken scaling, i.e., the rapidity distribution is
approximately given by (\ref{dNdy}) where the droplet rapidities $\eta_i$
are more or less evenly distributed between target and projectile rapidities.
Parametrizing the transverse and temporal distribution as gaussians,
we arrive at the droplet distribution
\bea
 \tilde{S}(x_i,K) \sim \exp\left[-\frac{m_\perp}{T}\cosh(Y-\eta_i) 
      -\frac{{\bf r}_{\perp,i}^2}{2R_A^2} - \frac{(\tau_i-\tau_f)^2}{2t_A^2} \right]
 \,, \label{SA}
\eea
where $\tau_i=\sqrt{t_i^2-z_i^2}$ is the invariant time and
$\tau_f$ the average freeze-out time.

The resulting correlation function becomes (in the system 
$Y=0$, where $\beta_l=\tanh Y=0$ and $\beta_o=p_\perp/m_\perp$)
\bea
 C_2(q) &=& 1+\exp[-q_s^2(R_A^2+R_d^2)             
        -q_o^2(R_A^2+R_d^2+\beta_o^2(t_A^2+t_d^2)) \nonumber \\
        &&  - q_l^2(\tau_f^2\frac{T}{m_\perp}+R_d^2
 )]     \,.      \label{CAy}
\eea
We observe that the larger nuclear size $R_A$ dominates the smaller
droplet size $R_d$ when droplets overlap.

\begin{figure}[t]
\psfig{figure=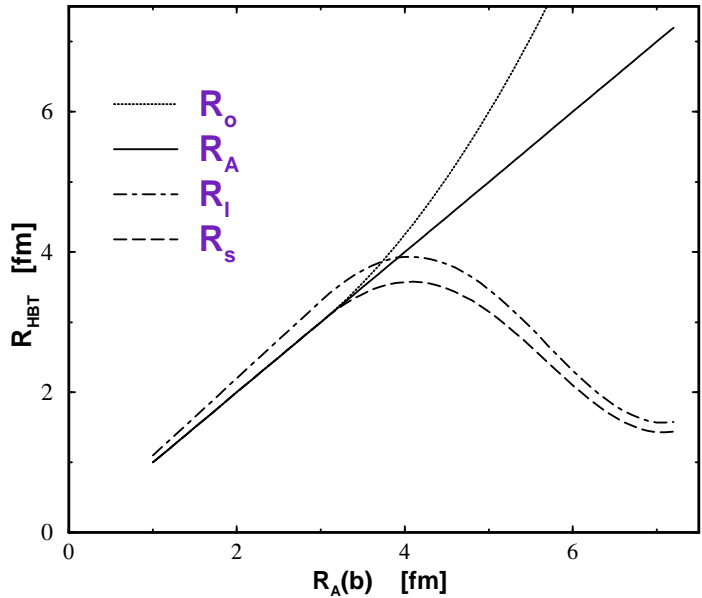,height=3.7in,angle=-90}
\caption{The HBT radii as function of nuclear overlap $R_A$ which is
proportional to centrality, $dN/dy$ or $E_\perp$. Droplet formation
is assumed to set from semicentral collisions. A long mixed phase will
then lead to large outward HBT radius $R_o$ whereas triggering on
large rapidity fluctuations corresponding to small droplets leads to
smaller longitudinal $R_l$ and sideward $R_s$ HBT radii.  \label{Fig2}}
\end{figure}

\section{Centrality dependence of HBT radii}

We can now study the consequences of forming droplets at RHIC
energies.  The onset of large rapidity fluctuations is one signal but
it should be accompagnied by the following behavior of the HBT radii.
In peripheral collisions, where the nuclear overlap and stopping is
small, we do not expect sufficient energy densities to build op and
form droplets. Thus the HBT radii should simply grow with the size of
the geometrical nuclear overlap, centrality, $dN/dy$ and transverse
energy $E_T$.  At SPS energies the HBT radii are indeed found to scale
approximately with the geometrical sizes of the colliding systems as
given by (\ref{CAy}).  

If energy densities achieved in RHIC collisions
are sufficient to form droplets in central collisions, the HBT
radii will deviate from the geometrical overlap if triggered on
large rapidity fluctuations (see Fig. 2)
as given by Eq.(\ref{Cdy}). Comparing the theoretical predictions of Eqs.
(\ref{Cdy}) and (\ref{CAy}) with the experimentally measured HBT radii
of Eq.(\ref{Cexp}) we see that the sideward and longitudinal HBT radii
decrease from the nuclear size $R_A,t_A$ to the droplet sizes
$R_d,t_d$. Thus at a certain semi-centrality, where energy densities
achieved in nuclear collisions 
start becoming large enough to create droplets,
the sideward and longitudinal HBT radii should bend over and start
decreasing with centrality.
The outward HBT radius may behave differently depending on
the duration of emission. The droplets may emit hadrons for a long
time, as is the case for a long lived mixed phase (referred to as the
``burning log''). In the hydrodynamic calculation of
Ref. \cite{Rischke}), the duration of emission and consequently the
outward HBT radius increase drastically up to five times larger than
the transverse size of the system, i.e. $R_A$.  This scenario is
indicated in Fig. 2.

Besides droplets we may expect some hadronic
background.  It is straight forward to include such one in the
correlation function.  As its spatial extend is expected to be on the
scale $\sim R_A$, it will reduce the correlation function at small
$q\sim\hbar/R_A$.  However, the droplets will still lead to
correlations at large relative momenta of order $q\sim\hbar/R_d$.
The large $q$ correlations are suppressed by the square of
the fraction of pions emerging from droplets at a given
rapidity.

\section{Summary}

 If first order transitions occur in high energy nuclear collisions,
density fluctuations are expected which may show up in rapidity
fluctuations event-by-event.  HBT interferometry adds an important
space-time picture to the purely momentum space information one gets
from single particle spectra.  Triggering on rapidity fluctuations, the
HBT radii may display a curious behavior. The outward HBT radius
$R_o$ may increase drastically with centrality due to a long lived
mixed phase. The longitudinal $R_l$ and sideward $R_s$ HBT radii will,
however, saturate and {\it decrease} for the very central collisions
because a large rapidity fluctuation signals a hot and dense droplet
of small size.

 The predicted behavior for the sideward and longitudinal HBT radii
is {\it opposite} to that predicted in cascade and hydrodynamic calculations.
It would be a clean signal of a first order phase transition in
nuclear collisions.

\end{document}